 \documentstyle[12pt]{article}

\newcommand{\beq}{\begin{equation}}
\newcommand{\eeq}{\end{equation}}
\begin{document}
\title
{The Isospin Splittings of Heavy-Light Quark Systems}

\author{Leonard S. Kisslinger and Zhenping Li\\
      Department of Physics, Carnegie Mellon University\\
      Pittsburgh, PA 15213}

\maketitle
\indent
\begin{abstract}
The isospin mass splittings of the pseudoscalar and vector D and B
light-heavy quark system have been calculated using the method of QCD sum
rules. Nonperturbative QCD effects are shown to be very small, so that
mass splittings arise almost completely from current quark mass splitting
and electromagnetic effects, for which a new gauge invariant QED treatment
is used.  The results are consistent with experiment.
A measurement of the isospin splitting of the
vector B mesons would give valuable information about quark mass splittings.
\vspace{3mm} \\

\end{abstract}
\newpage
\subsection*{1. Introduction}

The QCD sum rule method is well suited for the study of the mass splittings of
isospin multiplets.  In principle, this method allows a consistent treatment
using QCD and QED field theory of the three sources of isospin
symmetry breaking: the current quark mass differences in
the QCD Lagrangian, the nonperturbative QCD isospin violations which arise
from u-d flavor dependence of vacuum condensates, and electromagnetic
effects. In our first work on the isospin splittings of heavy-light
mesons\cite{kl}, however, we found that the standard treatment using
the two-loop diagrams
for electromagnetic corrections to order $\alpha_e$ was not gauge
invariant.  Recently we have developed a gauge-invariant theory for QED
corrections to the two-point functions used in QCD sum rules\cite{kl1},
which permits a consistent treatment of
all sources of isospin splitting within the QCD sum rule method.

In the present paper we obtain the mass splittings of the D and B
pseodoscalar and vector mesons.
Among the first applications of the method of QCD
sum rules was the study of isospin violations in the $\rho-\omega$
system\cite{svz} (see Ref.\cite{gl} for a review of the early work in this
area).  Recently, the technique has been used to
study the neutron-proton mass difference\cite{hhp,yhhk}, the octet baryon
mass splittings\cite{adi} and the mass differences in the charmed meson
systems (the D and $\rm{D}^*$ scalar and vector mesons)\cite{ei}.
In contrast with the phenomenological models of heavy-light meson
isospin violations\cite{ee,ch,gh}, the method of QCD sum rules allows one to
separate the current quark mass from the nonperturbative QCD effects.
Since the nonperturbative QCD isospin violations are smaller for the
heavy-light quark systems than for the light quark systems\cite{kl,kl2},
the isospin splittings in these charmed and bottom meson systems
should be given largely by quark mass differences and electromagnetic effects.
Moreover, since the electromagnetic effects and quark mass splittings
enter with different relative signs in the charmed vs the bottom systems,
the isospin splittings for a heavy-light quark system are particularly
promising for a determination of the current quark masses. (See Ref.\cite{gl}
for a discussion of early work on isospin violations and quark mass
differences.)

   The formulation of QED corections given in our previous
publication\cite{kl1}, in which we solved the gauge invariant problem by
including an additional current proportional to the total charge of the
charged current as required by the expansion of the link operator,
provides a consistent framework to calculate the gauge invariant two-point
functions in the operator product expansion.
Using this theory we find that the method is consistent with
experiment, which was not true in our earlier attempt\cite{kl} to use
phenomenological Coulomb potentials to estimate electromagnetic effects;
and that we can obtain a consistent QCD/QED sum rule analysis of the isospin
mass splittings of the D,$\rm{D}^*$, B and $\rm{B}^*$ systems.
This will be our main conclusion.

The starting point of the QCD sum rules is to evaluate the two point
function
\begin{equation}\label{1111}
\Pi_{\mu\nu} (q^2) = i \int d^4 x e^{iqx} < T(J_{\mu}(x) {\bar J}_{\nu}(0)
) >
\end{equation}
in the Wilson operator production expansion.  To investigate the violation
of the isospin symmetry,  the two point function  $\Pi(q^2)$ can be written
as
\beq\label{1}
\Pi(q^2)=\Pi_0(q^2)+m_q\Pi_m(q^2)+
 \Pi_{em}(q^2).
\eeq
$\Pi_0(q^2)$ in Eq. \ref{1} is  the leading term for the light heavy
quark system, the isospin violation comes from the isospin splitting of the
quark condensates in the non-perturbative power correction, which
has been given in previous publications\cite{kl,AE83}. The second
term in Eq. \ref{1} is the light quark mass, $m_q$, expansion of the
total correlator;  and the third term corresponds to the electromagnetic
corrections, which have been not treated consistently in the literature.

The paper is organized as follows. The light quark expansion correlator,
$\Pi_m(q^2)$, and
the electromagnetic effects, $\Pi_{em}$, are discussed in
Sections 2 and 3.  In Section
4, we shall present our numerical analysis of the isospin splittings for the
charm and bottom mesons, and finally the conclusion is given in Section 5.

\subsection*{2. The Contributions From The Up and Down Quark Mass Difference}
We write $\Pi_m(q^2)$ in Eq. \ref{1} as
\begin{equation}\label{2}
\Pi_m(q^2)=C_I I + C_3 \langle\bar q q\rangle
+C_5 \langle  \bar q(\sigma \cdot G) q\rangle,
\end{equation}
where the coefficient $C_I$  represents the perturbative contributions
and we limit the expansion to dimension = 5 operators.
Based on the results in Ref.
\cite{gb}, we find the $m_q$ expansion to order $\alpha_s$ is
\begin{equation}\label{3}
Im \left \{ C_I^{ps}\right \}=\frac {3M_Q}{4\pi} (1-x)\left [1+\frac
{4\alpha_s}
{3\pi}\left (f(x)+\frac 34x+\frac 32 x
\ln \left(\frac {x}{1-x}\right )\right )\right ]
\end{equation}
for the pseudoscalar and
\begin{equation}\label{4}
Im \left \{ C_m^{v}\right \}=Im \left \{C_m^{ps}\right \}
-\frac {\alpha_sM_Q}{\pi^2} (1-x)
\end{equation}
for the vector correlator, where $x=\frac {M^2_Q}{q^2}$,
\begin{equation}\label{5}
f(x)=\frac 94+2l(x)+\ln(x)\ln(1-x)+
\left (\frac 52-x-\frac 1{1-x}\right )\ln(x)-
\left (\frac 52-x\right )\ln(1-x)
\end{equation}
and $l(x)=-\int_0^x \ln(1-y)\frac {dy}{y}$ is the Spencer function.

The coefficients $C_3$ and $C_5$ in Eq. \ref{2} come from expansions in
the small quark mass of the light quark propagator. The relevant terms are
\begin{eqnarray}\label{51}
\langle q^{\alpha}(x){\bar q}^{\beta}(x) \rangle
=\delta^{\alpha\beta} \bigg ( \frac {i}{2\pi^2x^4}\hat{x} -\frac
{m_q}{4\pi^2x^2}-\frac 12 \langle {\bar q}q\rangle+\frac {im_q \langle
{\bar q}q \rangle}{48}\hat {x} \nonumber \\
+\frac {x^2}{192}\langle g_c {\bar q} \sigma\cdot G q\rangle -
\frac {im_q \langle g_c {\bar q} \sigma \cdot G q\rangle x^2}{2^7\cdot
3^2}\hat {x}\bigg )
\end{eqnarray}
where $\hat {x}\equiv \gamma^{\mu} x_{\mu}$.
The result to the order $\alpha_s$ for the coefficient $C_3$ is
\begin{eqnarray}\label{6}
C^{ps}_3 & =&\frac 12\frac 1{q^2(1-x)}\bigg \{\frac x{1-x}-1 \\ \nonumber
&+&\frac {2\alpha_s}{3\pi}\left [\frac {x}{(1-x)}\left (6+3\ln\left (
\frac {x}{x-1}\right )\right )
-x\left (3-(6+3x)\ln\left(
\frac x{x-1}\right )\right )\right ]\bigg \}
\end{eqnarray}
for the pseudoscalar and
\begin{eqnarray}\label{7}
C^{v}_3& = & \frac 12\frac 1{q^2(1-x)}\bigg \{\frac x{1-x} \\ \nonumber
&+&\frac {2\alpha_s}{3\pi}\left [\frac {x}{(1-x)}\left (4+3\ln\left (
\frac {x}{x-1}\right )\right )
-x\left (1-x\ln\left(
\frac x{x-1}\right )\right )+2\right ]\bigg \}
\end{eqnarray}
for the vector currents.  The $m_q$ expansions of the one loop
corrections to the quark condensate are finite after the
mass renormalization, in which the pole mass is used.

The coefficient $C_5$ in Eq. 2 is
\begin{equation}\label{8}
C_5^{ps}=\frac {x}{p^4(1-x)^3}\left(\frac {3x}{2(1-x)}-1\right )
\end{equation}
and
\begin{equation}\label{9}
C_5^{v}=\frac {1}{p^4(1-x)^2}\left ( \frac {3x^2}{2(1-x)^2}+\frac
{5x}{6(1-x)}-\frac 1{3}\right )
\end{equation}
for the pseudoscalar and vector currents, respectively.

\subsection*{3. The Electromagnetic Effects}

We write the electromagnetic corrections in analogue to Eq.
\ref{2};
\begin{equation}\label{10}
\Pi_{em}(q^2)=D_I I + D_3 \langle\bar q q\rangle
+D_5 \langle  \bar q(\sigma \cdot F) q\rangle ,
\end{equation}
where the first term is a two-photon-loop perturbative correlator,
the second term represents a photon-loop correction to the quark
condensate, and the third is a dimension 5 quark-photon condensate in
analogue to the quark gluon condensate in QCD. In reality, the quark-photon
condensate, $\langle  \bar q(\sigma \cdot F) q\rangle$, is not well
understood because of the gauge invariance problem, although it has been
used in the analysis of the proton and neutron mass difference\cite{yhhk}.
 Thus,
we introduce a new parameter $\delta$ by relating the quark-photon condensate
to the quark-gluon condensate;
\begin{equation}\label{131}
\langle  \bar q(\sigma \cdot F) q\rangle=-\delta
\langle  \bar q(\sigma \cdot G) q\rangle,
\end{equation}
where $F$ and $G$ correspond to photon and gluon fields
respectively.

The gauge invariant electromagnetic current\cite{kl1} at order $\alpha_e$
is
\begin{equation}\label{13}
J_{\mu}(x)={\bar q}(x) \gamma_\mu Q(x)+ ie_T {\bar q}(x) \gamma_\mu
\int_0^x A_{\nu}(y)dy^{\nu} Q(x),
\end{equation}
where the second term is a gauge fixing term for the charged currents,
and comes from the expansion of the link operator $exp(ie_T\int_x^y
A_{\nu}(z)dz^{\nu})$ which is being inserted between the heavy and light
quark fields to
make the operator product expansion in QED gauge invariant, and $e_T$ is
the total charge of the system.

Thus, the operator product expansion in QED for charge neutral currents
at order $\alpha_e$ should have the same form as that in QCD at order
$\alpha_s$ except that the
coupling constant $\frac {4\alpha_s}{3\pi}$ in QCD is replaced by $
\frac {e_Q^2\alpha_e}{\pi}$ in QED (note that for neutral currents
the charges for the light and
heavy quarks are equal with opposite sign).   By separating the gauge
dependent and independent terms for the leading current $J_{\mu}=
{\bar q}(x) \gamma_\mu Q(x)$, we have shown in
Ref. \cite{kl1} that it is convenient to write the operator product
expansion in QED into three  parts;
\begin{equation}\label{111}
D^{c}=D^{0}+D^{Q}+D^{q}.
\end{equation}
The first term, $D^{0}$, in Eq. \ref{111} is the gauge independent term for
the leading current $J_{\mu}={\bar q}(x) \gamma_\mu Q(x)$, thus, it also
has the same form as its counterpart in QCD except that the coupling
constant is replaced by $-\frac {e_Qe_q\alpha_e}{\pi}$ in this case.
The second and third terms in Eq. \ref{111} include the gauge dependent
terms of the leading current that only involves the self-energy diagrams of
the heavy and light quarks and the terms generated by the gauge fixing
current, $J^\prime_{\mu}=ie_T {\bar q}(x) \gamma_\mu \int_0^x A_{\nu}(y)
dy^{\nu} Q(x)$, in Eq. \ref{13}.

The advantage of this approach is that the first term in Eq. \ref{111}
and the correlators for the charge neutral currents have
been derived in QCD\cite{gb}; we have
\begin{equation}\label{11}
Im\left \{ D_I^{ps} \right \}=\frac {3e_Q^2 \alpha_e M_Q^2 }{8\pi^2}
(1-x)^2 f(x)
\end{equation}
and
\begin{eqnarray}\label{12}
Im \left \{ D_I^{v} \right \}  =  \frac {e_Q^2 \alpha_e q^2 }{8\pi^2}
(1-x)^2 \bigg [ (2+x)\left (1+f(x)\right )
 -  (3+x)(1-x)\ln \left ( \frac {x}{1-x}\right ) \nonumber  \\
-\frac {2x}{(1-x)^2}
\ln (x) -5-2x-\frac {2x}{1-x}\bigg ]
\end{eqnarray}
for the charge neutral pseudoscalar and vector currents respectively,
where $f(x)$ is given in Eq. \ref{4}.

For the charged current, the first term in Eq. \ref{111} is essentially
given by  Eqs. \ref{11} and \ref{12} for pseudoscalar and vector mesons,
and we found that $D^q$ in Eq. \ref{111} is zero since it is proportional
to the wavefunction renormalization for a zero mass particle in the Landau
gauge\cite{kl1}.  Therefore, the only term left in Eq. \ref{111} is $D^Q$,
and its imaginary part is shown to be\cite{kl1}
\begin{eqnarray}\label{15}
Im \left \{ D_I^Q\right \}^{ps}=\frac {9\alpha_e e_Qe_TM_Q^2}{16\pi^2}\bigg
[ (1-x)^2 \ln \left (\frac {x}{1-x}\right )-2x(1-x)-2\ln(x) \nonumber
\\ \frac 32(1-x)^2\bigg ]
\end{eqnarray}
and
\begin{eqnarray}\label{16}
Im\left \{ D_I^Q\right \}^{v} = \frac {3\alpha_e e_T e_Q}{16\pi^2}
\bigg [q^2(2+x)\left ((1-x)^2\ln \left (\frac x{1-x}\right )-x(1-x)-\ln(x)
\right ) \nonumber \\
-M_Q^2\left ( x(1-x)+\ln (x)+\frac 32(1-x)^2\right )\bigg ],
\end{eqnarray}
where $e_T=e_q+e_Q$ for charged systems.

Similarly, one can calculate the one-photon loop corrections to the
quark condensate.  The coefficients $D_3$ for the charge neutral
system are just an extension of the one loop corrections in QCD; therefore,
we have
\begin{equation}\label{17}
D_3^{ps}=\frac {e_Q^2\alpha_e}{2\pi}\frac {M_Q}{q^2(1-x)}
\left [ 1+\frac 32 \ln \left (\frac {M_Q^2}{\mu^2}\right ) -3(1-x)
\left (1+x\ln\left (\frac {1-x}x\right )\right )\right ]
\end{equation}
and
\begin{equation}\label{18}
D_3^{v}=\frac {e_Q^2\alpha_e}{2\pi}\frac {M_Q}{q^2(1-x)}
\left [ -1+\frac 32 \ln \left (\frac {M_Q^2}{\mu^2}\right )+(1-x)
\left (1+x\ln\left (\frac {1-x}x\right )\right )\right ]
\end{equation}
for the charged neutral pseudoscalar and vector currents respectively.

The coefficient $D_3$ for the charged currents follows from the same procedure
used in deriving the coefficient $D_I$.  The first term in Eq. \ref{111},
$D_3^0$ in this case, can be obtained from Eqs. \ref{17} and \ref{18}
by changing the coupling constant to $-e_Qe_q\alpha_e$. We find that
$D^Q_3$ becomes zero after the mass renormalization, in which the pole
mass is used.  This is because $D_3^Q$ only corresponds to the self-energy
diagram, and its wavefunction renormalization integral is canceled exactly
by the gauge fixing term, for which the physics is the same as that of
 the vanishing $D^q$ for the two-photon-loop diagram.    The term $D_3^q$
is found to be
\begin{equation}\label{21}
D_3^q=\frac {e_Te_q\alpha_e}{2\pi}\frac {M_Q}{q^2(1-x)}
\left [ 1+\frac 32 \ln \left (\frac {M_Q^2}{\mu^2}\right )+\frac 32(1+x)
\ln\left (\frac {1-x}x\right )\right ]
\end{equation}
for both pseudoscalar and vector currents.  It is interesting to note that
the term $D_3^0$, present in both charged and neutral mesons, has opposite
signs for the pseudoscalar and vector currents. It plays the same role as
the hyperfine interaction in the phenomenological quark models, while
$D^q_3$, present only in the charged current, could be interpreted as the
contribution from the photon cloud. This shows the subtle
correlation between the photon interaction and the light quark condensates.
In particular, the effects for the charged bottom mesons should be larger
than the charged charm mesons, as the factor $e_Te_q$ for B mesons
in Eq. \ref{21} is two times larger than that of the D mesons.

Similar to the quark-gluon condensate, one could obtain the quark-photon
condensate term in the fixed point gauge, and there is no contribution
from the gauge fixing term in Eq. \ref{13} to the coefficient $D_5$.
Therefore, the coefficient $D_5$ is a simple extension of the corresponding
QCD diagrams, and we have
\begin{equation}\label{211}
D_5^{ps}=\left (\begin{array}{c} e_Q^2 \\ e_q^2 \end{array} \right )\frac
{M_Q^3}{2(q^2-M_Q^2)^3}+\left ( \begin{array}{c} -e_Q^2 \\ e_Qe_q
\end{array} \right ) \frac {M_Q}{(q^2-M_Q^2)^2}
\end{equation}
and
\begin{equation}\label{22}
D_5^{v}=\left (\begin{array}{c} e_Q^2 \\ e_q^2 \end{array} \right )\frac
{M_Q^3}{2(q^2-M_Q^2)^3},
\end{equation}
where the top and bottom matrix elements correspond to the neutral and
charged currents respectively. Eq. \ref{22} and   the
first term in Eq. \ref{211}  come from the term in which a
photon is being emitted by the light quark itself, whose counterpart
in QCD is the
fifth term in Eq. \ref{51}.  They have the same effects as the light quark
mass difference in B and D mesons, and physically could be regarded as
the contributions from the electromagnetic effects to the light quark
masses. The second term in Eq. \ref{211}
comes from the photon being emitted from the heavy quark, and thus
has different signs for the charged and neutral systems.  The physics of
this term is the same as the hyperfine interaction in the phenomenological
quark  models, which generates the mass differences between the vector and
pseudoscalar states.
Therefore, the dominant electromagnetic effects should come from the
photon corrections to the quark condensate and the two-photon loop
corrections, while the quark-photon condensate term  is only responsible
for the difference between the vector and pseudoscalar states.

\subsection*{4. The Numerical Results}
Defining the quantity $\omega^2=M_Q^2-q^2$ for the Borel transformation,
where $\omega^2$ measures the off-shell effects of the heavy quarks,
we find
\begin{equation}\label{23}
f^2_p\frac {M^4_p}{M_Q^3}e^{-\frac {M^2_p
-M_Q^2}{\omega_B^2}}=\Pi_t^p(\omega_B^2)
\end{equation}
and
\begin{equation}\label{24}
f_v^2\frac {M_v^2}{M_Q}e^{-\frac {M_v^2-M_Q^2}{\omega_B^2}}
=\Pi_t^v(\omega^2_B)
\end{equation}
for the pseudoscalar and vector current. The correlator
$\Pi_t(\omega_B^2)$
can be divided into the perturbative and nonperturbative parts:
\begin{equation}\label{25}
\Pi_t(\omega_B^2)=\int_0^{s^0} Im C_I^t(\omega^2)e^{-\frac
{\omega^2}{\omega_B^2}}d\omega^2+\Pi_{np}(\omega_B^2),
\end{equation}
where the perturbative term is written in dispersion integral form,
and the nonperturbative correlator $\Pi_{np}(\omega_B^2)$ comes
from the standard Borel transformations of the terms associated
with the condensates.
The masses $M_p$ and
$M_v$ in Eqs. \ref{10} and \ref{11} can be obtained
from
\begin{equation}\label{26}
M^2=M_Q^2+\omega_B^4\frac {d\ln \left (\Pi_t(\omega_B^2)\right )
}{d\omega_B^2}.
\end{equation}
 The mass differences between different isospin states are obtained
by adjusting the parameters $s^0$ in Eq. \ref{25} to insure that
the mass difference is independent of Borel parameter $\omega_B^2$.
The range of $\omega_B^2$ is chosen in such way so that the
continuum and the nonperturbative contributions
are minimum.  By fitting the masses and decay constants we find
the working range for the Borel parameter $\omega_B^2$ are
$ 1.4 < \omega_B^2 < 2.4$ for D mesons, $0.9 < \omega_B^2 <1.9$ for $D^*$
mesons, $3.9 < \omega_B^2 <4.9 $ for B mesons, and $3.3 < \omega_B^2 < 4.3$
for $B^*$ mesons, where $\omega_B^2$ has a unit of GeV$^2$.

In order to fit both $D$ and $B$ systems in the same time, we also include
the dependence of the light quark mass and the quark condensate on the
factor,
\begin{equation}\label{277}
L^a=\left [ \ln \left (M_Q/\Lambda\right )/\ln \left (\mu/\Lambda
\right )\right ]^{a}
\end{equation}
where $a=-\frac 49$  for light quark masses and $\frac 49$ for the
quark condensate are the anomalous dimensions.
The quark condensate $\langle {\bar q}q\rangle$ has the standard value
$-0.0138$GeV$^3$, and the quark-gluon condensate is related to the quark
condensate by $\langle {\bar q}\sigma \cdot G q\rangle=m_0^2
\langle {\bar q}q\rangle$ with $m_0^2=0.8$ GeV$^2$.  As a first
approximation, we let $m_0^2$ be independent of the isospin splitting
so that the ratio of quark-gluon condensates between up and down quarks is
the same as that of the quark condensate, which is determined by the
parameter $\gamma$;
\begin{equation}\label{27}
\gamma=\frac {\langle \bar {d}d\rangle}{\langle \bar {u}u\rangle}-1.
\end{equation}

There are some constraints on the isospin differences of the light quark
masses and  condensates: the estimate from the chiral perturbation theory
gives  $m_d-m_u=3.0$ MeV\cite{gl}, while
the parameter $\gamma$ in various estimates
ranges from $-0.0079$\cite{yhhk} to $-0.002$\cite{adi,ei}.  However, the
parameter $\gamma$ in previous studies may not be reliable because the
electromagnetic effects have not been consistently included.  As a first
attempt, we set $m_d-m_u=4$ MeV, $\gamma=-0.002$, and $\delta=0$,
by adjusting the threshold parameter $s^0$ in Eq. \ref{25} to obtain a
constant in the working region of $\omega_B^2$,  the resulting isospin mass
splittings are
\begin{equation}\label{28}
M^{\pm}-M^0=\left \{ \begin{array}{r@{\quad\quad}l}
3.7 & \mbox{ for D} \\ 2.9 &\mbox{ for
$D^*$}
\end{array}\right .
\end{equation}
and
\begin{equation}\label{29}
M^{\pm}-M^0=\left \{ \begin{array}{r@{\quad\quad}l}
-1.3 & \mbox{ for B} \\ -0.8 &\mbox{ for
$B^*$}
\end{array}\right .
\end{equation}
in units of MeV.  The contributions from the electromagnetic effects to
the isospin mass splittings can be obtained from the difference
between the isospin mass splittings with and without the electromagnetic
effects, and results are approximately
\begin{equation}\label{30}
\left \{ M^{\pm}-M^0\right \}_{em}=0.4
\end{equation}
for both $D$ and $D^*$ mesons and
\begin{equation}\label{31}
\left \{ M^{\pm}-M^0 \right \}_{em}=1.9
\end{equation}
for both $B$ and $B^*$ mesons.
Thus, the contributions from the electromagnetic effects are considerable
larger in B mesons. The dominant terms in the electromagnetic contributions
are the photon loop corrections to the quark condensate, and
the quark condensate for the
B mesons is also considerable larger than that for D mesons due to the
factor $L^{a}$ in Eq. \ref{277} in addition to the electric charge factor
in Eq. \ref{21}.

It is interesting to consider the phenomenological values
for the electromagnetic corrections obtained in previous work on
isospin splitting of the heavy-light mesons. In the quark model
estimate of Ref. \cite{ee}, in which the
$\left \{ M^{\pm}-M^0 \right \}_{em}$ is taken as the
Coulomb potential, the values for the $D$ and $B$ systems are approximately
equal.  This is not satisfactory, since $\left \{ M^{\pm}-M^0 \right \}_{em}$
 must be much larger
for the $B$ systems than the $D$ systems to agree with experiment.
In the work of Ref.\cite{gh} $\left \{ M^{\pm}-M^0 \right \}_{em}$ for
the $D$ systems is found
to be 7/5 times that of the $B$, causing an even larger discrepancy
with experiment if used by us.

We find little dependence of the isospin mass splitting on the parameter
$\gamma$; varying the parameter $\gamma$ from $-0.002$ to $-0.004$ only
generates a $0.1$ MeV increase of the mass splitting for $D$ mesons, and
no significant increase at all for $B$ mesons.  However, this may be due
to the assumption that the ratio of the quark gluon condensates has the
same dependence on the parameter $\gamma$ as that of
the quark condensates between the up and down quarks.  This shows that
the contributions from the quark gluon condensate is important, and the
heavy-light quark systems may not be a good place to determined the
parameter $\gamma$.

We show our best fit to the isospin mass splittings in Fig. 1 for $D$,
Fig. 2 for $D^*$, Fig. 3 for $B$, and Fig. 4 for $B^*$ states.
The parameters for this fit are $m_d-m_u=3.0$ MeV, $\delta=0.01$, and
$\gamma=-0.002$.  The resulting isospin splittings are
\begin{equation}\label{288}
M^{\pm}-M^0=\left \{ \begin{array}{r@{\quad\quad}l}
4.3 & \mbox{ for D} \\ 3.5 &\mbox{ for
$D^*$}
\end{array}\right .
\end{equation}
and
\begin{equation}\label{299}
M^{\pm}-M^0=\left \{ \begin{array}{r@{\quad\quad}l}
-1.2 & \mbox{ for B} \\ -1.1 &\mbox{ for
$B^*$.}
\end{array}\right .
\end{equation}
The corresponding differences of the threshold parameters $s$
are
\begin{equation}\label{388}
s^{\pm}-s^0=\left \{ \begin{array}{r@{\quad\quad}l}
0.012 & \mbox{ for D} \\ 0.026 &\mbox{ for
$D^*$}
\end{array}\right .
\end{equation}
and
\begin{equation}\label{399}
s^{\pm}-s^0=\left \{ \begin{array}{r@{\quad\quad}l}
-0.024 & \mbox{ for B} \\ -0.026 &\mbox{ for
$B^*$}
\end{array}\right .
\end{equation}
in units of GeV$^2$, which indeed have the same signs as those of
the light quark mass difference\cite{ei}.
This is in good agreement with data
($M_{D^{\pm}}-M_{D^0}=4.7 \pm 0.7$ MeV\cite{pdg} and
$M_{D^{*\pm}}-M_{D^{*0}}=3.32 \pm 0.13$ MeV\cite{bo92}
for D mesons and $-0.1 \pm 0.8$ MeV\cite{pdg} for $B$ mesons).
Perhaps more encouraging is that the light quark mass difference
in this fit is in good agreement with the result of the chiral perturbation
theory\cite{gl}.

An important feature of this investigation
is the differential vector-pseudoscalar isospin mass splittings
$\Delta (M)$;
\begin{equation}\label{36}
\Delta (M)= (M^{*\pm}-M^{*0})-(M^{\pm}-M^0),
\end{equation}
which provides very important probe of the hyperfine
splittings of the heavy-light quark potential.
The quantity $\Delta (M)$ comes from two sources: the hyperfine
interaction from the differences of the light quark masses and condensates
and from the electromagnetic effects (in particular, the quark-photon
condensate in Eq. \ref{211}). These two sources have the same sign for the
$D$ mesons and opposite signs for $B$ mesons. Therefore, the quantity
$\Delta (M)$ for $B$ mesons really depends on the competition between these
two sources.  From Eqs. \ref{288} and \ref{299}, we have
\begin{equation}\label{37}
\Delta(M)=\left \{ \begin{array}{r@{\quad\quad}l}
-0.8& \mbox{ for D mesons} \\ 0.1 &\mbox{ for B mesons}
\end{array}\right .
\end{equation}
This is in good agreement with recent studies by Cutkosky
and Geiger\cite{cg}, the quantity $\Delta(M)$ for $B$ mesons is
essential zero within the theoretical uncertainty.
A measurement of the isospin mass splittings for the vector B mesons will
provide a crucial test in this regard.
A study in chiral perturbation
theory finds\cite{jenkins} that the contributions
from quark mass difference only gives 0.3 MeV for
D mesons, and the rest might comes from
the electromagnetic corrections.  Our calculation
is consistent with this conclusion, and furthermore
we also find that there is only a small contribution from
the difference of the quark condensates to $\Delta(M)$.

\subsection*{5. Summary}

Gauge invariance dictates that there is an additional gauge
fixing current for charged mesons.  This gauge fixing current
generates additional contributions to the electromagnetic effects for
charged mesons.  In particular, we find the photon-loop correction to the
quark condensate dominant for electromagnetic effects, and because of
this correlation between quark condensates and photons, the
electromagnetic corrections are considerably larger in B mesons than those
in D mesons.  This is quite contrary to the conclusions of phenomenological
quark models.  Indeed, our results are in good agreement with
the known data, and further experimental studies of $B^*$ mesons will
certainly provide us more insights into the hyperfine interactions
in the quark-quark potential.

We find little dependence of the isospin mass splitting on the isospin
differences of the quark condensates, a nonperturbative QCD effect given
by the parameter $\gamma$ in our approach,
while the light quark mass difference used in
our calculation is in good agreement with the result of the chiral
perturbation theory.  Therefore, a consistent treatment of the
electromagnetic effects would allow us to study the light quark mass
difference in heavy-light quark systems.

Another interesting application of our approach is the isospin splittings
in Kaon systems. Since the strange quark has much less mass than the charm
and bottom quarks, the mechanism of the electromagnetic effects might be
different from the heavy light quark systems.  This investigation is in
progress, and the result will be published elsewhere.

This work is supported by National Science Foundation grant
PHY-9319641.

\vspace{5mm}
\noindent {\Large\bf Figure Captions}

\vspace{5mm}
\begin{itemize}
\begin{enumerate}
\item The sum rule for $D$ mass splittings: $M^{\pm}-M^0$, with parameters
$m_d-m_u=3.0$ MeV, $\delta=0.01$ and $\gamma=-0.002$. The difference of the
threshold parameter is $s^{\pm}-s^0=0.012$ GeV$^2$ for the center curve,
$0.016$ and $0.008$ GeV$^2$ for the upper and lower curves repectively.
\item The same as Fig. 1 for $D^*$ mass splittings with the threshold
parameter difference $0.026$ for the center curve, $0.03$ and $0.022$
GeV$^2$ for the upper and lower curves respectively.
\item The same as Fig. 1 for $B$ mass splittings with the threshold
parameter difference $0.024$ for the center curve, $0.032$ and $0.016$
GeV$^2$ for the upper and lower curves respectively.
\item The same as Fig. 1 for $B^*$ mass splittings with the threshold
parameter difference $0.026$ for the center curve, $0.034$ and $0.018$
GeV$^2$ for the upper and lower curves respectively.
\end{enumerate}
\end{itemize}
\end{document}